\documentclass[a4paper]{emulateapj} 

\usepackage{hyperref}
\usepackage{color}

\hypersetup{
    bookmarks=true,         % show bookmarks bar?
    unicode=false,          % non-Latin characters in Acrobat’s bookmarks
    pdftoolbar=true,        % show Acrobat’s toolbar?
    pdfmenubar=true,        % show Acrobat’s menu?
    pdffitwindow=true,     % window fit to page when opened
    pdfstartview={FitH},    % fits the width of the page to the window
    pdftitle={field blue straggler stars},    % title
    pdfauthor={rsantucci},     % author
    pdfsubject={Astronomy},   % subject of the document
    pdfcreator={dvipdf},   % creator of the document
    pdfproducer={dvipdf}, % producer of the document
    pdfkeywords={blue straggler stars},% list of keywords
    pdfnewwindow=true,      % links in new window
    colorlinks=true,       % false: boxed links; true: colored links
    linkcolor=red,          % color of internal links
    citecolor=blue,        % color of links to bibliography
    filecolor=magenta,      % color of file links
    urlcolor=cyan,           % color of external links
    breaklinks=true,
    linktocpage
}

\newcommand{\metal}{[Fe/{}H]}

\newcommand{\teff}{$T_{\rm eff}$\,}
\newcommand{\logg}{log\,$g$\,}

\shorttitle{The Frequency of Field Blue Straggler Stars}
\shortauthors{Santucci et al.}

\begin{document}

\title{The Frequency of Field Blue-Straggler Stars in the Thick Disk and Halo System of the Galaxy}

\author{Rafael M. Santucci}
\affil{Departamento de Astronomia - Instituto de Astronomia,
Geof\'isica e Ci\^encias Atmosf\'ericas, Universidade de S\~ao Paulo,
S\~ao Paulo, SP 05508-900, Brazil}

\author{Vinicius M. Placco}
\affil{Gemini Observatory, Hilo, HI 96720, USA}

\author{Silvia Rossi}
\affil{Departamento de Astronomia - Instituto de Astronomia,
Geof\'isica e Ci\^encias Atmosf\'ericas, Universidade de S\~ao Paulo,
S\~ao Paulo, SP 05508-900, Brazil}

\author{Timothy C. Beers}
\affil{Department of Physics and JINA Center for the Evolution of the Elements, 
University of Notre Dame, 225 Nieuwland Science Hall, Notre Dame, IN 46556, USA}

\author{Henrique M. Reggiani}
\affil{Departamento de Astronomia - Instituto de Astronomia,
Geof\'isica e Ci\^encias Atmosf\'ericas, Universidade de S\~ao Paulo,
S\~ao Paulo, SP 05508-900, Brazil}

\author{Young Sun Lee}
\affil{Department of Astronomy and Space Science, Chungnam National 
University, Daejeon 305-764, Republic of Korea}

\author{Xiang-Xiang Xue}
\affil{Max-Planck-Institute for Astronomy K\"{o}nigstuhl 17,
D-69117, Heidelberg, Germany, and \\ 
Key Lab of Optical Astronomy, National Astronomical
Observatories, CAS, 20A Datun Road, Chaoyang District, 100012,
Beijing, China}

\author{Daniela Carollo}
\affil{INAF - Osservatorio Astrofisico di Torino,
Via Osservatorio 20, Pino Torinese, 10020 Torino, Italy}

\begin{abstract}

We present an analysis of a new, large sample of field blue-straggler stars
(BSSs) in the thick disk and halo system of the Galaxy, based on
stellar spectra obtained during the Sloan Digital Sky Survey (SDSS) and
the Sloan Extension for Galactic Understanding and Exploration (SEGUE).
Using estimates of stellar atmospheric parameters obtained from application of
the SEGUE Stellar Parameter Pipeline, we obtain a sample of
some 8000 BSSs, which are considered along with a previously selected
sample of some 4800 blue horizontal-branch (BHB) stars. We derive the
ratio of BSSs to BHB stars, F$_{\rm BSS/BHB}$, as a function of
Galactocentric distance and distance from the Galactic plane. The
maximum value found for F$_{\rm BSS/BHB}$ is $\sim~$4.0 in the thick
disk (at 3 kpc $<$ $|$Z$|$ $<$ 4 kpc), declining to on the order of
$\sim~1.5-2.0$ in the inner-halo region; this ratio continues to decline
to $\sim~$1.0 in the outer-halo region. We associate a minority of field
BSSs with a likely extragalactic origin; at least 5$\%$ of the BSS
sample exhibit radial velocities, positions, and distances commensurate
with membership in the Sagittarius Stream.

\end{abstract}


\section{Introduction} 
\label{c1}

Blue-straggler stars (BSSs) lie brighter and blueward of the main-sequence 
turnoff region in color-magnitude diagrams of globular and open clusters. 
This is a region where, if the BSSs are single stars, they should have
already evolved away from the main sequence. Since they lie blueward of
the cluster turnoff point, and appear to linger (or straggle behind
already evolved stars), they have been traditionally referred to as blue
stragglers \citep{Sandage53}. BSSs have been found in all stellar populations 
and environments, including: Population I stars in the field \citep{Carney05}, 
in open clusters of all ages \citep{deMarchi06}, and among thick-disk stars 
\citep{Fuhrmann99}, as well as among Population II stars in globular clusters 
\citep{Piotto04}, the Galactic halo \citep{Preston94}, the Galactic bulge 
\citep{Clarkson11}, and dwarf galaxies in the Local Group \citep{Momany07}.

Since their discovery, a large number of studies and reviews have been
published discussing the detection of BSSs, their properties, and
various hypotheses to account for the blue-straggler phenomenon
\citep{Leonard89,Stryker93, Bailyn95,Leonard96,Preston14}. Two mechanisms 
for their origin have been most prominently mentioned: (i) Collisions
between stars, resulting in a more massive and apparently younger star
\citep{Hills76}; and (ii) coalescence and/or mass-transfer processes in
binary systems, initially proposed by \citet{McCrea64}. The latter
scenario appears to be the main channel for the formation of BSSs, where
the more-massive star in a binary system transfers material, during its
post main-sequence stage, to its presently-observed companion
\citep{Boffin14}. The companion star, now more massive, appears
anomalously younger.

\citet{FusiPecci92} were among the first to propose that BSSs in different
environments may have different origins. In globular clusters, both of
the aforementioned processes could act simultaneously. Collisions are
expected to dominate in the central regions, while in the outer regions,
BSSs mass transfer dominates. As shown by \citet{Mapelli04}, the specific 
frequency of BSSs in globular clusters appears highly peaked at cluster 
centers, and rapidly decreases at intermediate radii; it rises again in 
the outer regions of most globular clusters.

A constrasting trend is found when comparing the frequency of BSSs in
different environments \citep{Preston94,Momany07}; denser stellar
environments (globular clusters) exhibit BSS frequencies that are lower
than in less-dense stellar environments (open clusters and the Galactic
field). One explanation for this observation suggests that multiple
systems (presumably the progenitors of BSSs) can be more easily
disrupted when subjected to gravitational interactions within the
clusters \citep{Preston00}. Even the stars found in the low-density
outer regions of globular clusters may be subject to stellar encounters
if their intra-cluster motions carry them into the central high-density
regions.

\citet{Preston00} argued that field BSSs were created primarily by 
mass-transfer processes. They noted that field BSSs are much more common per 
unit luminosity than BSSs in the lower-density outer regions of globular 
clusters. Binary systems, even in those low-density regions, may have 
already been destroyed, which would reduce the number of BSSs.
\citet{Preston00} obtained high-resolution spectroscopy for
62 of the 84 so-called blue metal-poor (BMP) stars brighter than $B \sim
15$ from \citet{Preston94}. These stars exhibit main-sequence gravities
and colors bluer than most metal-poor globular cluster turnoffs,
$0.15~<~(B-V)_0~<~0.35$. \citet{Preston00} concluded that over
60$\%$ of their sample are binaries, and that at least 50$\%$ of them
are BSSs. These values suggest that the BSSs can be a strong indicator
of the presence of binary systems, and can be used as probe of the
binarity (in particular close binaries) of the stellar populations in
the Galaxy.

\citet{Glaspey94} obtained moderate-resolution spectra of the Li 6707 \AA~
doublet for a sample of likely field BSSs, and obtained clear 
evidence for Li depletion, concluding that this depletion is a general 
property associated with blue stragglers. These authors argued that the 
origin of the Li depletion in their sample stars was likely related to 
large-scale mixing, presumably associated with the formation of blue 
stragglers.

\citet{Ryan01} and \citet{Ryan02} have also argued in favor of the binarity
of field blue stragglers, based on consideration of the so-called
ultra-Li-deficient halo stars, proposing that these stars may have had
their surface abundances of lithium reduced by the same mechanism that
produces field blue stragglers. These ultra-Li-deficient halo stars were
shown to possess more rapid rotation than other ancient metal-poor
stars, with the spin-up likely due to the mass-transfer process, and/or
chemical peculiarities associated with mass transfer from an AGB
progenitor. The Ryan et al. sample of stars were cooler than the
main-sequence turnoff, hence they were referred to by these authors as
``blue stragglers to be.'' These stars thus could be viewed as the
low-mass counterpart of field blue stragglers \citep{Ryan02}.
Furthermore, the process of mass transfer that appears to be responsible
for creating BSSs must act in cooler stars as well. \citet{Rochap01} 
identify cooler blue-straggler candidates (yellow or red stragglers). 
These were referred to as {\it chromospherically young and kinematically 
old stars} (CYKOS).

It is worth noting that the BSS frequency in low-luminosity dwarf
galaxies and open clusters \citep[${\rm N_{BSS}/N_{BHB}}$\footnote{N$_{\rm BSS}$ 
and N$_{\rm BHB}$ are the number of BSSs and BHB stars found in these environments,
respectively.}$\sim 2.2$;][]{Momany07} is different than that derived
for the Galactic halo \citep[${\rm N_{BSS}/N_{BHB}}$ $\sim$ 4;][]{Preston94}. 
However, as pointed out by \citet{Momany07}, the latter value was found based 
on a sample of only 62 BMP stars distributed along different lines of sight, 
located at different distances, and for which no observational star-by-star 
(BSS/BHB) correspondence could be defined.

Furthermore, field-star samples have introduced another puzzle, one at
least as intriguing as the blue straggler phenomenon itself.
\citet{Preston94}, \citet{Preston00} and \citet{Carney01,Carney05} 
argued that binary-star evolution and mass transfer is one path for the
creation of blue stragglers, and apparently the most common among field
stars. However, some metal-poor main-sequence field stars that are
hotter than globular-cluster main-sequence turnoffs are apparently {\it
not binaries} \citep[see][for further details]{Preston94}. These
authors accounted for their existence as the debris from accreted dwarf
satellite galaxies whose star formation continued over an extended
period.

This paper aims to examine the issues previously raised concerning field
BSSs by gathering a large new sample of BSS candidates, in order to
improve estimates of the specific frequency of N$_{\rm BSS}/N_{\rm BHB}$
(F$_{\rm BSS/BHB}$) in the Galactic field, and to test the hypothesis
of the extragalactic origin for some BSSs. This paper is outlined as
follows. Section~\ref{c2} describes the assembly of the database used to 
identify our sample of field BSSs and BHB stars, and the photometric 
restrictions adopted for both samples. Section~\ref{c3} summarizes our 
analysis of the selected spectra, and the techniques used to partition them
into these two groups. Section~\ref{c4} derives estimates for the absolute 
magnitudes and distances for both BSSs and BHB stars, and evaluates their 
frequencies in the Galactic field. Section~\ref{c5} presents a kinematic 
analysis of the field BSS sample, in an attempt to associate their radial 
velocities with stellar streams. A brief discussion and our conclusions 
are provided in Section~\ref{c6}.

\section{Photometric Selection and Stellar Parameters} 
\label{c2}

\subsection{SDSS -- DR8}

The Sloan Digital Sky Survey \citep[SDSS;][]{York00} is a large
photometric and spectroscopic survey that covers about one-quarter of the
northern sky. This paper makes use of the eighth data release (DR8),
which contains the data obtained by the survey through January 2011
\citep{Aihara11}. The first restrictions adopted in the selection of
field BSSs and BHB stars follow the same criteria used by
\citet{Xue08}.

\subsection{Color Cuts}

Blue-straggler and blue horizontal-branch candidates are often
identified in the SDSS color-color diagram region with the following
color cuts: $0.60 < (u-g) < 1.60$ and $-0.50 < (g-r) < 0.05$
\citep{Xue08,Deason11b}. This region is preferentially populated by A-type
stars. For this work, we first obtained a sample of 19771 objects within
these color limits, and with available spectra, but with no restrictions
on apparent magnitudes or location on the sky within the SDSS footprint,
and covering the range $14.5 < g < 21$, the bright limit being set by
saturation in the SDSS imaging scans. Note that the
$ugriz$ magnitudes and color indices are all corrected for absorption
and reddening using the extinction maps from \citet{Schlegel98}.

Previous efforts to study stars in this color region usually targeted
only BHB stars \citep{Sirko04,Xue08,Deason11b,Xue11}. In this work, we
are interested in selecting both BSS and BHB candidates. However,
photometry alone does not uniquely distinguish main-sequence A-type stars (of
which the metal-poor stars are BSS candidates) from giant (BHB)
candidates. Furthermore, in this color region, the A-type stars can be
mixed with stars of cooler spectral types, such as metal-poor F- and
G-type stars. Thus, we also made use of available spectroscopy to trim
our sample, as described below.

\subsection{Stellar Parameters}

The spectra of the selected sample have stellar parameters determined by
the current version of the SEGUE Stellar Parameter Pipeline \citep[SSPP,
][]{Lee08a,Lee08b,Allende08,Smolinski11,Lee11}. From the SSPP, we
obtained values of effective temperature (\teff), surface gravity
(\logg), metallicity (\metal{\footnote{The abundance of an element in
relation to another is defined in the usual manner, using the notation:
{[A/{}B]} = $log(N_A/{}N_B) _{\star} - log(N_A/{}N_B)_{\odot}$, where
$N_A$ and $N_B$ are the number of atoms of elements A and B in the star
($\star$) and in the Sun ($\odot$), respectively.}}), and heliocentric
radial velocity ($V_{\rm HRV}$). To select the best BSS candidates (as
opposed to foreground A-type stars in the disk populations), we applied
a restriction in metallicity, \metal~$<-0.4$, reducing the sample to
18560 objects.

It is relatively straightforward to identify BSSs in globular clusters
because the horizontal-branch stars, whose colors blend with hotter
main-sequence stars, are seen at different apparent magnitudes 
\citep[BHB stars are typically $\sim 2$ magnitudes brighter than BSSs, as noted 
by][]{Preston94,Deason12}. Unfortunately, these two populations cannot be 
easily separated in the field, because the stars are not at the same distance.

\begin{figure} 
\epsscale{1.22} 
\plotone{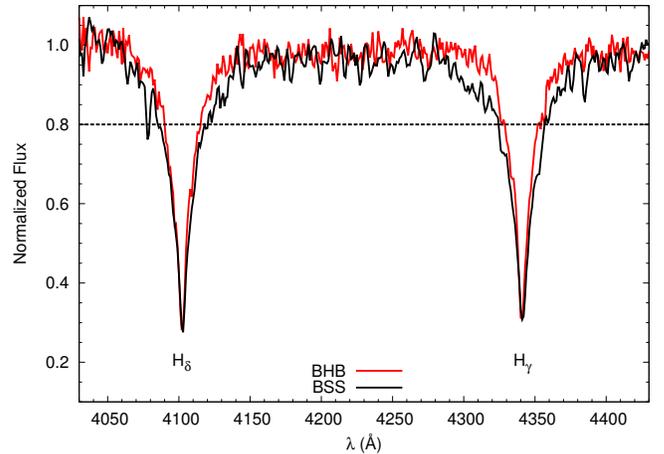}
\caption{Normalized spectrum of a BHB star (red line) and a BSS (black 
line) for the same \teff ($\sim$8100 K) in the H$_{\gamma}$-H$_{\delta}$
region. The dashed line is drawn at 20$\%$ below the continuum,
highlighting the difference between the Balmer-line wings. \\}
\label{bssXbhb}
\end{figure}

In the temperature range of A-type stars, the Balmer lines are the main
tool to separate A-type main-sequence from giant stars (BSSs from BHB
stars, respectively). The temperature modifies the depth of the Balmer
lines, while the surface gravity changes the width of its wings
\citep[as initially noted by][]{Pier83}. These effects have a clear
impact on the appearance of the spectrum, as shown in
Figure~\ref{bssXbhb}: for stars with the same \teff~(in this case,
$\sim$8100 K), the Balmer-line wings are broader for higher surface gravities. 

\section{The Blue-Straggler and Blue Horizontal-Branch Samples} 
\label{c3}

In this section we describe the methods used to separate BSSs from BHB
stars, using the available medium-resolution SDSS/SEGUE spectroscopy.
\citet{Deason12} states that the differences between BSSs and BHB stars
becomes most evident for spectra with S/N$>$5 \citep[see the bottom panels in
Figure 3 of][]{Deason12}. For this work, we prefer to only use stars
with spectra having an average signal-noise ratio (over the optical
spectrum) of $\langle$S$/$N$\rangle$ $\geq$ 10. This limit allows for a
confident distinction between BSSs and BHB stars where they are
photometrically indistinguishable, and for the reliable stellar parameters 
from the SSPP.

Using the parameters calculated by the SSPP, we can compare the surface
gravities and effective temperatures of BSSs with BHB stars
\citep[similar to][]{Wilhelm99b, Deason12}. However, the atmospheric
parameters calculated for stars with temperatures above 7500~K (where
the BSSs and BHB stars are usually found) require additional quantitative
validation. This is accomplished by comparing the results of the SSPP
with other methods for separating the BSSs from BHB stars, which
generally required higher S/N spectra, typically greater than $\sim$15
(values typically achieved for SDSS spectra of stars with apparent
magnitudes $g~<~18$).

It is also important to note that, even though stars with $g~>~18$ are
usually not suitable for analysis by all the methods presented below,
they provide valuable information on the populations of BSSs and BHB
stars (especially since, as shown below, the majority of likely BSSs in
our sample are fainter than $g = 17$). For the following analysis, we
divided the sample in two: (i) The Bright Sample ($g~<~18$) and (ii) the
Faint Sample ($g~\geq~18$).

\subsection{The Bright Sample: $14.5~<~g~<~18$}

Since the Bright Sample contains stars having spectra with higher S$/$N
ratios, it is possible to apply a series of quantitative tests to
distinguish between BSSs and BHB stars. These are described below.

\subsubsection{SSPP Restrictions}

\begin{figure} 
\epsscale{1.22} 
\plotone{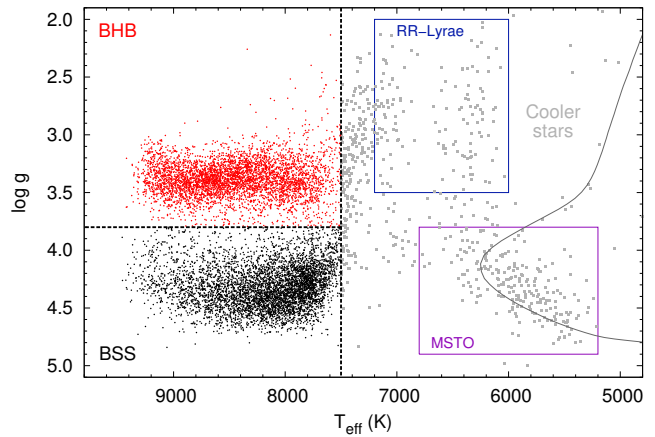}
\caption{Distribution of the SSPP \teff vs. \logg for stars with $g$ $<$
18. There are two stellar groups concentrated in \teff $\geq$ 7500 K: one
for \logg $\sim$ 3.4 (BHB stars: red dots) and other with \logg $\sim$ 4.3
(BSSs: black dots). The gray squares are cooler stars and are not considered
further in our analysis. For convenience of the reader, we highlight the positions
where the RR Lyrae stars are often found using the blue rectangle
\citep{Wilhelm99b}, where the gap with no stars present is the expected 
location of the instability strip; the purple rectangle shows the region 
where the occupied by main-sequence turnoff stars (MSTO). Overplotted is a 
Yale-Yonsei isochrone \citep{Demarque04} for 12 Gyr and \metal$=-2.0$. \\} 
\label{tempxlogg_goodmMag_limits} 
\end{figure}

The parameters provided by the SSPP were used to identify BSSs and BHB
stars. Since the Balmer lines are strong in the 7500~K $<$ \teff $<$
10000~K range, it is possible to effectively distinguish between BSSs
and BHB stars from their \logg distribution \citep{Wilhelm99b}.
Figure~\ref{tempxlogg_goodmMag_limits} summarizes the adopted
restrictions. The line separating these two groups of A-type stars is
\logg $=$ 3.80, as estimated from the \logg~distributions, for stars
with $g~<~18$ (see top panel of Figure~\ref{histo_logG_mag}). This value
represents a threshold at which, statistically, the contamination of the
BSS population by BHB stars is minimal. The limits lie at 3$\sigma$ from
the peaks in the top panel of Figure~\ref{histo_logG_mag} centered at
\logg = 3.38 ($\sigma$ $=$ 0.14 dex) and \logg = 4.33 ($\sigma$ $=$ 0.18
dex), which represent the BHB stars and BSSs, respectively.

\subsubsection{The f$_{m}$ vs. D$_{0.2}$ Method}

The parameters used in the following analysis are obtained through fits
to a Sersic profile \citep[][]{Sersic68}, which describes the shape of
the Balmer spectral features \citep{Clewley02}:

\begin{equation} 
S(x)=n-a \cdot {\rm{exp}}[-(|\lambda-\lambda_0|/b)^c],
\label {eq1} 
\end{equation} 

\noindent where, $n$, $a$, $b$, and $c$ are free parameters. The parameter $n$
fits the continuum level ($\sim$1 for our normalized spectra),
$\lambda_{0}$ is the wavelength at the line center, and $a$ defines the
scale-depth parameter. The parameters $b$ (scale width) and $c$ (scale
shape) will be subsequently examined. They are used in the scale
width-shape method \citep{Clewley02}, as described in Section~\ref{cb}.

There is no fundamental relation between $a$ and the other Sersic
parameters, but the $a$ values have well-defined Gaussian distributions
around a mean value for each Balmer line. For our sample, these average
values are: $\langle$a$_{\beta}\rangle$ $=$ 0.709,
$\langle$a$_{\gamma}\rangle$ $=$ 0.694, and $\langle$a$_{\delta}\rangle$
$=$ 0.769. We refined the estimated Sersic profile parameters by
re-adjusting the function with $\langle${\it{a}}$\rangle$ fixed for each
line.

The f$_m$ versus D$_{0.2}$ method was first proposed by \citet{Pier83};
it compares \teff and \logg indirectly, by measuring specific quantities
in the stellar spectra. The parameter f$_{m}$ is the deepest flux
relative to the continuum level, usually measured at the center of the
line (it is sensitive to \teff). The parameter D$_{0.2}$ is the width of
the line measured at 20$\%$ below the adjusted continuum level
\citep{Beers92,Sirko04}, and provides an indicator of \logg. As seen in
Figure~\ref{bssXbhb}, when \logg{} is higher, the greater D$_{0.2}$
becomes. To obtain D$_{0.2}$, we subtracted the roots of the inverse
function when $S(x) = 0.8n$, following \citet{Clewley02}, which yields:

\begin{equation}
D_{0.2}=2 \cdot b \cdot [{\rm{log}}(n/(5 \cdot \langle{a}\rangle)]^{\frac{1}{c}}.
\label{eq2} 
\end{equation}

\begin{figure} 
\epsscale{1.22} 
\plotone{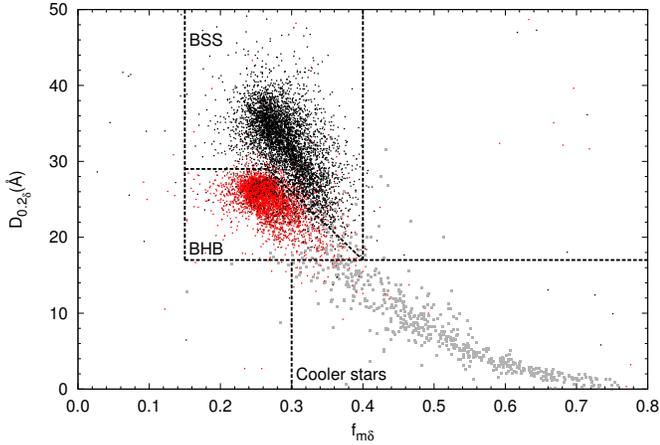}
\caption{Comparison of f$_{\rm m\delta}$ vs. D$_{0.2\delta}$, color-coded 
according to the restrictions adopted by the SSPP
(Figure~\ref{tempxlogg_goodmMag_limits}). The distributions of BSSs and
BHB stars exhibit some overlap beyond the adopted limits (dashed lines). The
adopted criteria to select BSSs are: (i) 0.15 $<$ f$_{\rm m\delta}$
$<$ 0.27, D$_{0.2\delta}$ $>$ 29; (ii) 0.27 $<$ f$_{\rm m\delta}$ $<$
0.40, D$_{0.2\delta}$ $>$ $-$90 $\cdot$ f$_{\rm m\delta}$ $+$ 53.
Accordingly, the criteria for BHB candidates are: (i) 0.15 $<$ f$_{\rm
m\delta}$ $<$ 0.27 and 17 $<$ D$_{0.2\delta}$ $<$ 29; (ii) 0.27 $<$
f$_{\rm m\delta}$ $<$ 0.40 and 17 $<$ D$_{0.2\delta}$ $<$ $-$90
$\cdot$ f$_{\rm m\delta}$ $+$ 53. \\} 
\label{fmxD02deltapipelinegoodMagregions} 
\end{figure}

We adopted the H$_{\delta}$ feature to compare the parameters f$_{\rm
m}$ and D$_{0.2}$ for this method, indicated by $\delta$ in the
subscript of the derived parameters \citep[see][for further
details]{Sirko04}. Figure~\ref{fmxD02deltapipelinegoodMagregions} shows
the behavior of $D_{0.2\delta}$, as a function of f$_{\rm m\delta}$, for
stars with $g < 18$. The stars concentrated at ($f_{\rm m\delta};
D_{0.2\delta})\sim(0.26;25)$ are the BHB stars; the stars with larger
$D_{0.2\delta}$ are BSSs. The remaining stars have cooler spectral
types. 

To compare the restrictions adopted by the SSPP with the limits in the
f$_{\rm m}$ vs. D$_{0.2}$ method, we identified the stars in
Figure~\ref{fmxD02deltapipelinegoodMagregions} according to the regions
described in Figure~\ref{tempxlogg_goodmMag_limits}; black dots
represent BSSs, red dots are BHB stars, and the gray squares are cooler stars.
Note that these divisions exhibit good agreement with the separation based on the
SSPP stellar-parameter values ($\sim$92\% of stars satisfy both restrictions).

\subsubsection{The Scale Width-Shape Method: {\it{c}} vs. {\it{b}}} 
\label{cb}

\begin{figure} 
\epsscale{1.22} 
\plotone{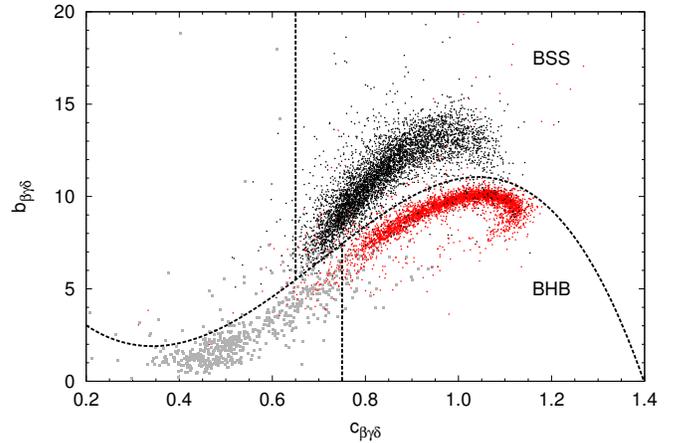}
\caption{Comparison between $c_{\beta\gamma\delta}$ and
$b_{\beta\gamma\delta}$, with color-coding according to the
restrictions adopted by the SSPP. As verified with the $f_{\rm m}$ vs.
$D_{0.2}$ method, the limits adopted by the SSPP for separating the
BSSs and BHB stars agree very well ($\sim$95\% of stars satisfy both 
restrictions). The limit line that separates these two groups is 
defined by the fourth-order polynomial given by Equation~\ref{eqarray}. \\} 
\label{cxb_pipeline_goodMag_adjust} 
\end{figure}

An additional method often adopted to separate BSSs from BHB stars is
the scale width-shape method ($c$ vs. $b$), first described by
\citet{Clewley02}. This method separates these stars through the
parameters set by the Sersic profile (with the parameter $a$ fixed). In
order to apply this approach, we calculated the average of the
parameters $c$ and $b$ for the H$_{\beta}$, H$_{\gamma}$ and
H$_{\delta}$ lines, following \citet{Deason11b}. The average of these
parameters is shown in Figure~\ref{cxb_pipeline_goodMag_adjust}, where
$c_{\beta\gamma\delta}$ and $b_{\beta\gamma\delta}$ are color-coded by
the adopted regions described in Figure~\ref{tempxlogg_goodmMag_limits}.
As can be seen, there is good agreement when comparing this method with
separation based on the SSPP stellar-parameter values ($\sim$95\% of
stars satisfy both restrictions). We note that the parameters $b$ and $c$ 
could be used, in principle, by themselves to separate BSS from BHB 
\citep[see][]{Preston14}. The mixed stars that passed the
initial color cuts now exhibit a clear separation in the
$c_{\beta\gamma\delta}$ versus $b_{\beta\gamma\delta}$ plane; the gap
between them distinguishes between BSSs and BHB stars, as defined
by a fourth-order polynomial:

\begin{eqnarray} 
b_{\beta\gamma\delta}= +9.20 - 46.32 \cdot
(c_{\beta\gamma\delta}) + 82.24 \cdot 
(c_{\beta\gamma\delta})^2 \nonumber \\ 
- 23.36 \cdot (c_{\beta\gamma\delta})^3 -10.82 
\cdot (c_{\beta\gamma\delta})^4.
\label{eqarray} 
\end{eqnarray}

To select BSSs with $g~<~18$, we combined the SSPP restrictions, $f_{\rm
m}$ vs. $D_{0.2}$, and the scale width-shape method. Whereas the
constraints of the SSPP proved fully compatible with the methods based
on the analysis of the Balmer lines, we further restrict the BSS and BHB
candidates according to the limits: BSSs have \logg $>$3.8 and the BHB
stars lie between 3.0$~\leq~$\logg$~\leq~$ 3.8. The mean error on
effective temperature given by the SSPP is about 150~K \citep{Lee08b},
which is slightly higher than the average error of the parameters for
our sample ($\sim$130 K). We selected only the stars with \teff $\geq
7650$~K. The stars satisfying these cuts are within 3$\sigma$~of the
mean of both peaks in \logg, as shown in the top panel of
Figure~\ref{histo_logG_mag}, resulting in 4838 BSSs and 4380 BHB stars.

\subsection{The Faint Sample: $18~<~g~<~21$} 

The traditional methods ($f_{\rm m}$ vs. $D_{0.2}$, and $c$ vs. $b$)
cannot properly evaluate the stars in our sample with $g > 18$, due to
the lower $\langle$S$/$N$\rangle$ ratios in SDSS/SEGUE spectra of these
fainter stars \citep{Clewley02,Sirko04}. Thus, for the 5663 stars in
the Faint Sample, we employed only the SSPP parameters to separate the
BSSs and BHB stars.

\begin{figure} 
\epsscale{1.22} 
\plotone{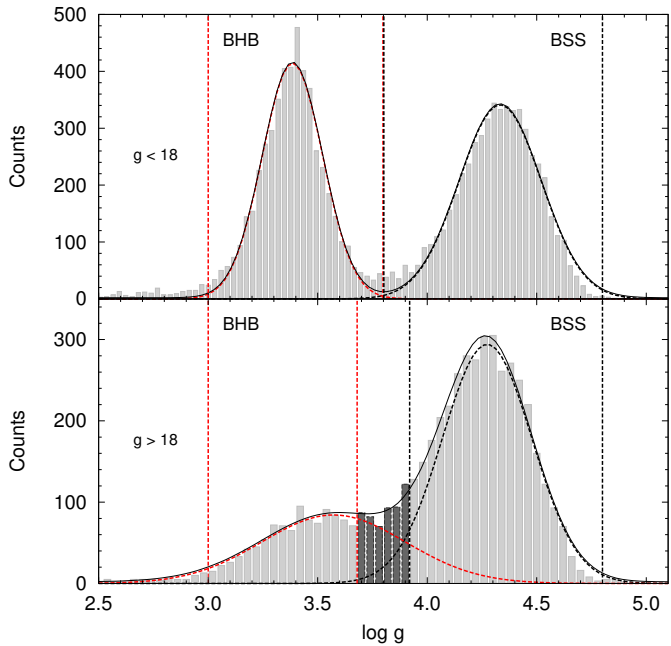} 
\caption{Top panel: Distribution of \logg for stars with $g < 18$. The peak
centered at \logg $=$ 3.38 ($\sigma$ $=$ 0.14 dex) represents the
distribution of BHB stars; that at \logg $=$ 4.33 ($\sigma$ $=$ 0.18
dex) applies to BSSs. Bottom panel: Distribution of \logg for stars with
$g > 18$. The peak centered at \logg $=$ 3.57 ($\sigma$ $=$ 0.33 dex)
represents the distribution of BHB stars; that at \logg $=$ 4.27
($\sigma$ $=$ 0.20 dex) represents BSSs. It can be seen that the
fraction of BHB stars decays quickly at high surface gravity for the
fainter stars. \\}
\label{histo_logG_mag}
\end{figure}

The \logg distribution for the Faint Sample is shown in the lower panel
of Figure~\ref{histo_logG_mag}. The peak centered at \logg $=$ 3.57,
which represents BHB stars, has a high dispersion ($\sigma=0.33$~dex).
This increases the size of the overlap region between the two
distributions and, if we were to consider the 3$\sigma$ region as a
limit, it would exclude almost all BSSs, which are peaked at \logg $=$
4.27 ($\sigma=0.20$~dex).

As the stars analyzed in the lower panel of Figure~\ref{histo_logG_mag}
are also well-described by Gaussian distributions, we assessed the
probability of finding BHBs using the peak positions and values up to
3$\sigma$ away from them. The adopted restriction for the BHB
distribution is \logg $<$ 3.66 (3$\sigma$ away from the BSSs peak).

Since our goal is to select BSSs candidates, we adopt a more relaxed
criteria to its distribution, even though it leads to greater
contamination from the BHB population. Estimating the amount of stars
coming from each distribution, we analyzed the contamination for every
sigma. For example, there are 3646 stars with \logg $>$ 4.27 (located at 
\logg$_{\rm {BHB}}$ $+$ 2$\sigma_{\rm {BHB}}$), but the probability of 
finding BHB stars in this range of \logg is approximately 2\%. 
Thus, stars with \logg $>$ 4.27 have a 98\% probability of being
BSSs.

We conclude that, for $g~>~18$, the most reliable BHB candidates are
found in the surface-gravity range 3.00 $<$ \logg $<$ 3.66 (416
objects). Accordingly, for BSSs we adopted the limit of \logg $>$ 3.92, 
which is 1$\sigma$ away from the BHB peak, with an average probability of
$\sim$88$\%$.

Combining the Bright and Faint Samples, we obtained a total 
of 8001 BSS candidates, with an average probability of $\sim$95$\%$. 
We also obtained 4796 BHB candidates, with an average probability of 
correct selection of $\sim$98$\%$.

\begin{figure}
\epsscale{1.22}
\plotone{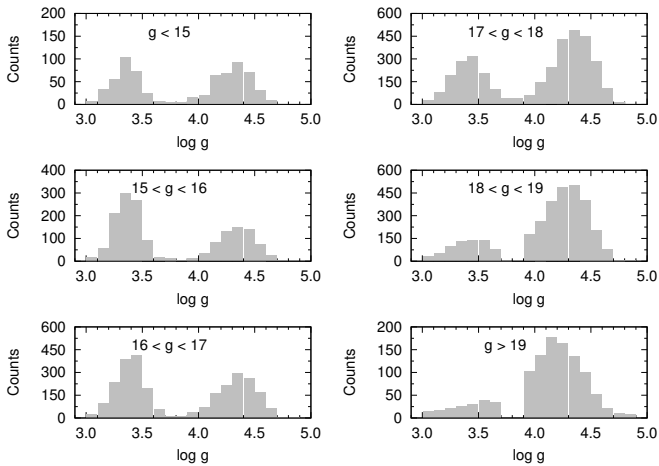}
\caption{Histograms of \logg,  evaluated in different intervals of
$g$-band apparent magnitude. 
The number of BSSs dominates over the number of BHB stars at fainter apparent
magnitude ($g >$ 17). \\}
\label{histo_mag_logg}
\end{figure}

\begin{figure}
\epsscale{1.22}
\plotone{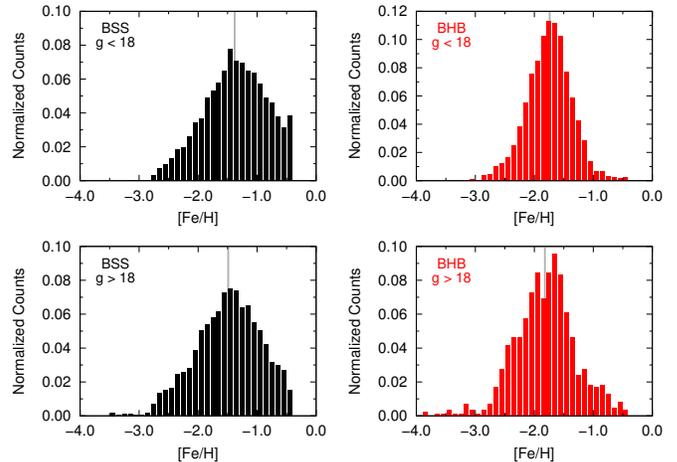}
\caption{\metal~distributions for our sample stars (left panels: BSSs;
right panels: BHB stars) for the Bright Sample (upper panels) and the
Faint Sample (bottom panels). The gray lines indicate the mean of each
\metal~distribution. \\}
\label{MDFs}
\end{figure}

Figure~\ref{histo_mag_logg} shows the distributions of \logg from the
SSPP for our full sample of BSSs and BHB stars, in intervals of apparent
magnitude. As is clear from inspection of this sample, the higher
gravity BSSs begin to dominate over the lower gravity BHB stars
for $g > 17$. Figure~\ref{MDFs} shows the distribution of derived
\metal~for these same stars, in this case for the Bright Samples
and Faint Samples of BSSs and BHB stars. Although the mean metallicities
of the BSSs and BHB stars are similar, the BSSs are more prevalent at higher
metallicity than the BHB stars.

\section{Distance Calibrations and Frequencies} 
\label{c4}

In this section we obtain an absolute magnitude calibration for our full
sample of 8001 BSSs and 4796 BHB stars, extending over the
magnitude range $14.5 < g < 21$.

A number of studies adopt calibrations for apparent magnitudes based on
photometric systems other than the SDSS $ugriz$ system \citep[e.g.,
][]{Kinman94,Preston94,Preston00,Carney01,Carney05}. In this work, we
used the transformations of \citet{Zhao06}, which were derived from SDSS
stars with known $UBVRI$ photometry, including a number of BSSs and
BHB stars:

\begin{equation} 
{\rm V_0} = g - 0.561 \cdot (g-r) - 0.004, 
\label{eqV}
\end{equation} 

\noindent and 

\begin{equation} (B-V)_{0} = 0.916 \cdot (g-r)+ 0.187.
\label{eqBV} 
\end{equation}

According to \citet{Beers12}, the valid range for these conversions is 
$-0.5 < g-r < 1.0$, consistent with the color indices adopted in this work.
In addition, the calculated values are within the limits proposed by
\citet{Carney05} to define field BSSs.

\subsection{Absolute Magnitudes of Field BSSs}

Based on the $V$-band magnitude, it is possible to estimate the 
absolute magnitude for BSSs using the relation from \citet{Kinman94}: 

\begin{equation} 
{\rm {M_{V_{BSS}}}}= 1.32 + 4.05 \cdot (B-V)_{0} - 0.45 \cdot {\rm {\metal}}.
\label{eqkin} 
\end{equation}

This relation was first determined from BSSs in globular clusters
by the work of \citet{Sarajedini93,Sarajedini94}. Assuming $(B-V)_{0} = 0.1$
and \metal~$=-1.7$ as typical values for metal-poor A-type stars
selected in this work, we verify that the field BSS candidates have $M_V
\approx 2.5$. This value is in agreement with blue stragglers observed
in globular clusters and dwarf galaxies \citep{Sarajedini93,Momany07}.

\citet{Deason11b} also developed a calibration for the $g$-band absolute
magnitude for field BSSs. They made use of stars in Stripe 82 that are
members of the Sagittarius Stream, combining the apparent magnitudes and
the results obtained by \citet{Watkins09}. The latter used RR Lyr$\ae{}$
stars to estimate the distance to stars in the stream in the right
ascension range 25$^{\rm {o}} < \alpha < 40^{\rm {o}}$, with D$_{{\rm {Sgr}}}
= 26.1\pm5.6$ kpc. The absolute-magnitude calibration in the $g$-band
from \citet{Watkins09}, M$_{\rm g_{BSS}} = 3.108 + 5.495
\cdot (g - r)$, is similar to that presented by \citet{Kinman94} for
\metal~$=$ $-$1.5, and also agrees with the mean metallicity of stars
identified in the Sagittarius Stream \citep[\metal~$=-$1.43,
][]{Watkins09}. However, unlike the relation used by \citet{Kinman94},
\citet{Deason11b} found no strong dependence on metallicity. This is
probably due to the use of BSSs in the Sagittarius Stream, which already
have a mean \metal~ = $-$1.4, and also have similar luminosities
\citep{Watkins09}. The adopted error for this calibration is
$\sigma_{{\rm {M_{g}}}}~=~$0.5 mag, primarily arising from the expected
error in the distance to the Sagittarius Stream.

The distances are systematically smaller for the method calibrated to
BSSs in the Sagittarius Stream, but both relations agree within
2$\sigma$. We adopt the calibration given in Equation~\ref{eqkin} of
\citet{Kinman94}, because it was determined from BSSs in globular
clusters with different metallicities covering a wide range of colors
and absolute magnitudes. Furthermore, we take into account that the
error in the absolute magnitude of each star is certainly greater for
the fainter objects.

\subsection{Absolute Magnitudes of Field BHBs}

We employed two different calibrations to determine absolute magnitudes
for the BHB sample: (1) \citet{Deason11b} derived M$_{\rm g}$ using SDSS
photometry of 10 globular clusters from \citet{An08}; (2)
\citet{Fermani13} proposed a calibration with a dependence on the
metallicity, arguing that the relation adopted by \citet{Deason11b} is
systematically fainter for metal-poor stars and brighter for metal-rich
stars. The absolute magnitudes for \citealt{Deason11b} (M$_{\rm gD}$)
and \citealt{Fermani13} (M$_{\rm gFS}$) are given by:

\begin{eqnarray} 
{\rm {M_{gD}}}= 0.434 - 0.169 \cdot (g - r) + 2.319 \cdot {(g- r)}^2 \nonumber \\ 
+ 20.449 \cdot {(g - r)}^3 + 94.517 \cdot {(g - r)}^4, 
\label{eqdes} 
\end{eqnarray} 

\begin{eqnarray} 
{\rm {M_{gFS}}}= 0.0075 \cdot e^{-14.0 \cdot (g - r)} \nonumber \\ 
+ 0.04 \cdot ({\rm {\metal}} + 3.5)^2 + 0.25.
\label{eqfer} 
\end{eqnarray}

The range of M$_{\rm g}$ for BHB stars in globular clusters evaluated by
\citet{Deason11b} is $0.45 <$ M$_{\rm g} < 0.65$, and their metallicities are in
the range [$-2.3:-1.3$]. The authors did not find any significant dependence 
on metallicity for M$_{\rm g}$.

According to \citet{Fermani13}, the lack of metallicity dependence on
the absolute magnitude calibration of \citet{Deason11b} is primarily due
to uncertainties in their adopted distance moduli, reddening, and
presumed systematic differences between clusters and field halo stars;
they thus argued that Deason et al. could not rule out a detectable
metallicity dependence on field BHB stars. For the distance calculations
presented below, we explore both M$_{\rm g}$ calibrations.

\subsection{The Relative Frequencies of BSSs and BHB Stars}

\subsubsection{Galactocentric Distances}

In order to determine the ratio between BSSs and BHB stars (F$_{\rm BSS/BHB}$),
we first proceed by counting them directly, as a function of their distance
from the Galactic center. The Galactocentric distance, $R$, is
calculated based on the distance, $D$, from the Sun (in kpc):

\begin{equation}
D = 10^{\left[\left({\rm{m-M}}\right)/5-2\right]},
\label{dist1} 
\end{equation}

\noindent and

\begin{equation} 
R^2 = (R_{\odot}-D\cos{b}\cos{l})^2 + 
(D\sin{b})^2 \nonumber \\ 
+ (D\cos{b}\sin{l})^2, 
\label{rg}
\end{equation} 

\noindent where $b$ and $l$ are the Galactic latitude and longitude,
respectively, and $R_{\odot}$ is the distance of the Sun to the Galactic
center. We adopted $R_{\odot}$ $=$ 8.5 kpc to be consistent with the
recent kinematic analysis from \citet{Carollo10}.

\begin{figure} 
\epsscale{1.22} 
\plotone{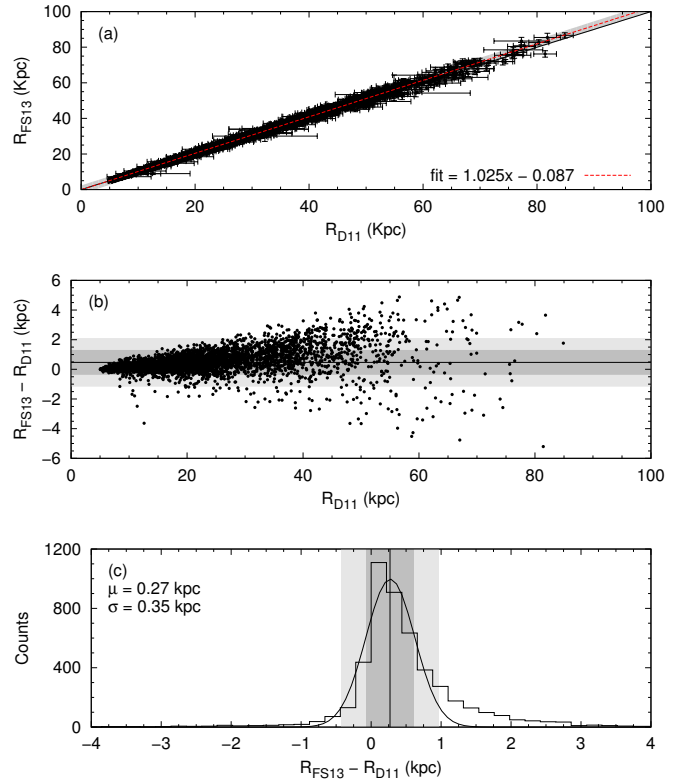} 
\caption{(a): Comparison between distances estimated using the absolute
magnitude calibrations of \citet{Deason11b} and \citet{Fermani13}. The solid
line represents $R_{\rm FS13}$ $=$ $R_{\rm D11}$, the red dashed line is a linear
fit to the data, and the gray region is the $\pm$ 3$\sigma$ limit of the data.
Panels (b) and (c) show, respectively, the residuals for the distribution and
its dispersion, represented by a Gaussian fit. The dark gray region represents
the limit within 1$\sigma$ of the average value ($\sim$0.34 kpc) for the
distribution; the light gray region is the limit within 2$\sigma$ of the 
average value. \\} 
\label{Distance_Compare} 
\end{figure}

Figure~\ref{Distance_Compare} compares the Galactocentric distances obtained
using the BHB calibrations from \citet{Deason11b} ($R_{\rm D11}$) and
\citet{Fermani13} ($R_{\rm FS13}$). The $R_{\rm FS13}$ distances are, on
average, $\sim2.5\%$ larger than $R_{\rm D11}$, which is within the errors for
both calibrations. Also, from panels {\it {b}} and {\it {c}} of
Figure~\ref{Distance_Compare}, it is possible to see that, even though there is
a larger spread for distances above 40 kpc, there are no trends between the
calibrations. The zero-point offset for this comparison is 0.27 kpc, with a
standard deviation of 0.39 kpc.

\begin{figure} 
\epsscale{1.22} 
\plotone{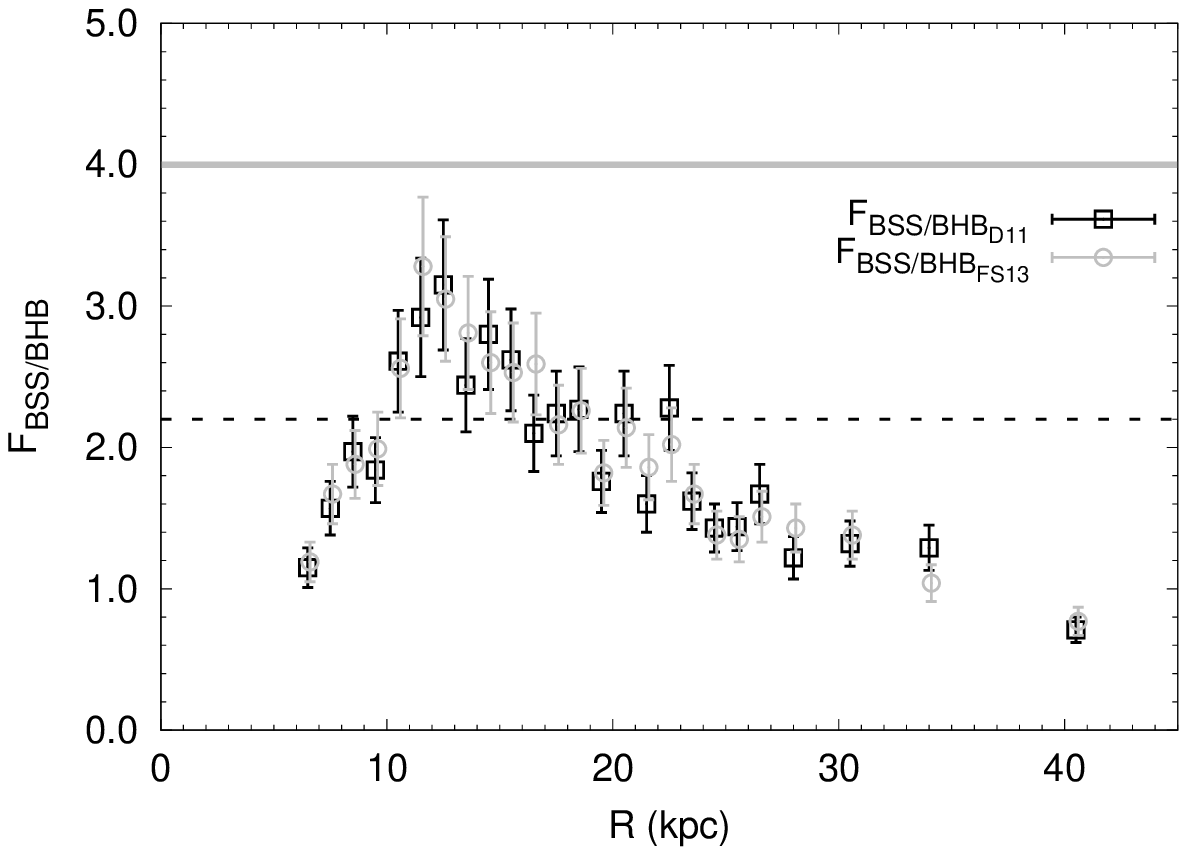}
\caption{Comparison of the ``as observed'' F$_{\rm BSS/BHB}$, as a function of 
Galactocentric distance, using the absolute-magnitude calibrations from
\citet{Deason11b} and \citet{Fermani13}. The gray line represents the
value of F$_{\rm BSS/BHB}$ estimated in the Solar Neighbourhood by
\citet{Preston94}, and the dashed line represents the value of F$_{\rm
BSS/BHB}$ found in nearby dwarf galaxies by \citet{Momany07}. Corrections 
for the sampling biases of these populations are described in the text. \\}
\label{fbsxbhb_Rerrors} 
\end{figure}

It is clear that the differences between these two approaches increase
for larger distances. For this reason, we compared the ``as observed''
frequency F$_{\rm BSS/BHB}$ for both R$_{\rm D11}$ and R$_{\rm FS13}$,
to gauge the possible impact of choosing one or the other of these
calibrations. Results are shown in Figure~\ref{fbsxbhb_Rerrors} (values
can be found in Table~\ref{tablefreq}). Both frequencies agree within
3$\sigma$, and the residual values have an average of zero, with a
standard deviation of 0.16 kpc. Thus, we choose to employ the
calibration from \citet{Deason11b} to represent the distances to BHB
stars in the Galactic field, since we found that the introduction of the
\metal~as a parameter does not have any impact on the derived distances.

\begin{figure}
\epsscale{1.22}
\plotone{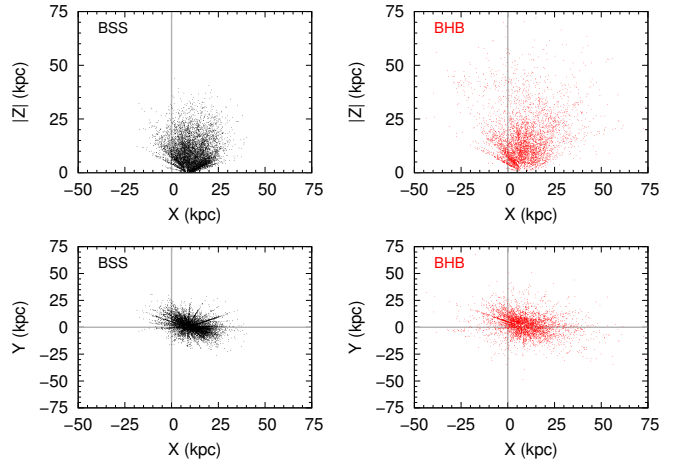}
\caption{Distribution of our sample stars (left panels: BSSs; right
panels: BHB stars) in the XZ plane (upper panels), and in the XY plane
(the Galactic plane; lower panels). The gray lines are drawn for
reference to include the location of the Galactic center. \\}
\label{XYZ_Dist}
\end{figure}

For completeness, Figure~\ref{XYZ_Dist} shows the derived distances of
the BSSs and BHB stars in our sample plotted in the XYZ planes of the Galaxy. As
expected, the BHB stars explore to greater distances than the BSSs (and
are undersampled in the Solar Neighborhood relative to the BSSs, due to
the SDSS bright limit), so care must be taken to account for these
limitations in obtaining our final derived relative frequency estimates. 

\begin{center} 
\begin{deluxetable*}{ccccccccc} 
\tabletypesize{\scriptsize}
\tablecolumns{9} 
\tablewidth{0pc} 
\tablecaption{Stellar Counts in Galactocentric Distance Intervals\label{tablefreq}}
\tablehead{&&
\multicolumn{3}{c}{\citet{Deason11b}}&
\colhead{}&
\multicolumn{3}{c}{\citet{Fermani13}}\\ 
\cline{3-5} \cline{7-9}\\ 
\colhead{R(kpc) }&
\colhead{N$_{\rm BSS}$ }& 
\colhead{N$_{\rm BHB}$ }& 
\colhead{F$_{\rm BSS/BHB}$ }& 
\colhead{$\sigma_{F_{\rm BSS/BHB}}$ }& 
\colhead{ }& 
\colhead{N$_{\rm BHB}$ }&
\colhead{F$_{\rm BSS/BHB}$ }& 
\colhead{$\sigma_{F_{\rm BSS/BHB}}$ }} 
\startdata

06$-$07  &  155  &   99  &  1.57  &  $\pm$0.14  &&   93  &  1.67  &  $\pm$0.14  \\
07$-$08  &  242  &  123  &  1.97  &  $\pm$0.19  &&  129  &  1.88  &  $\pm$0.21  \\
08$-$09  &  293  &  159  &  1.84  &  $\pm$0.25  &&  147  &  1.99  &  $\pm$0.24  \\
09$-$10  &  366  &  140  &  2.61  &  $\pm$0.23  &&  143  &  2.56  &  $\pm$0.26  \\
10$-$11  &  426  &  146  &  2.92  &  $\pm$0.36  &&  130  &  3.28  &  $\pm$0.35  \\
11$-$12  &  501  &  159  &  3.15  &  $\pm$0.42  &&  164  &  3.05  &  $\pm$0.49  \\
12$-$13  &  508  &  208  &  2.44  &  $\pm$0.46  &&  181  &  2.81  &  $\pm$0.44  \\
13$-$14  &  470  &  168  &  2.80  &  $\pm$0.33  &&  181  &  2.60  &  $\pm$0.40  \\
14$-$15  &  483  &  184  &  2.63  &  $\pm$0.39  &&  191  &  2.53  &  $\pm$0.36  \\
15$-$16  &  463  &  220  &  2.10  &  $\pm$0.36  &&  179  &  2.59  &  $\pm$0.35  \\
16$-$17  &  437  &  195  &  2.24  &  $\pm$0.27  &&  202  &  2.16  &  $\pm$0.36  \\
17$-$18  &  465  &  205  &  2.27  &  $\pm$0.30  &&  206  &  2.26  &  $\pm$0.28  \\
18$-$19  &  354  &  201  &  1.76  &  $\pm$0.30  &&  195  &  1.82  &  $\pm$0.30  \\
19$-$20  &  386  &  172  &  2.24  &  $\pm$0.22  &&  180  &  2.14  &  $\pm$0.23  \\
20$-$21  &  318  &  199  &  1.60  &  $\pm$0.30  &&  171  &  1.86  &  $\pm$0.28  \\
21$-$22  &  335  &  147  &  2.28  &  $\pm$0.20  &&  166  &  2.02  &  $\pm$0.23  \\
22$-$23  &  271  &  167  &  1.62  &  $\pm$0.30  &&  162  &  1.67  &  $\pm$0.26  \\
23$-$24  &  207  &  145  &  1.43  &  $\pm$0.20  &&  150  &  1.38  &  $\pm$0.21  \\
24$-$25  &  194  &  135  &  1.44  &  $\pm$0.17  &&  144  &  1.35  &  $\pm$0.17  \\
25$-$26  &  192  &  115  &  1.67  &  $\pm$0.17  &&  127  &  1.51  &  $\pm$0.16  \\
26$-$27  &  140  &  115  &  1.22  &  $\pm$0.21  &&   98  &  1.43  &  $\pm$0.18  \\
27$-$29  &  229  &  173  &  1.32  &  $\pm$0.15  &&  166  &  1.38  &  $\pm$0.17  \\
29$-$32  &  225  &  174  &  1.29  &  $\pm$0.16  &&  216  &  1.04  &  $\pm$0.17  \\
32$-$36  &  166  &  235  &  0.71  &  $\pm$0.16  &&  215  &  0.77  &  $\pm$0.13  \\
36$-$45  &   89  &  407  &  0.22  &  $\pm$0.09  &&  413  &  0.22  &  $\pm$0.10  \\
 Total   & 7915  & 4391  &  1.80  &  $\pm$0.22  && 4349  &  1.82  &  $\pm$0.23

\enddata 
\tablecomments{There are 76 BSSs and $\sim$65 BHB stars within R $<$ 6 kpc. 
There are also 10 BSSs and $\sim$360 BHB stars that lie between 
45 kpc $<$ R $<$ 100 kpc.}
\end{deluxetable*} 
\end{center}

\subsubsection{Refined Relative Frequency Estimates}

\begin{figure} 
\epsscale{1.22} 
\plotone{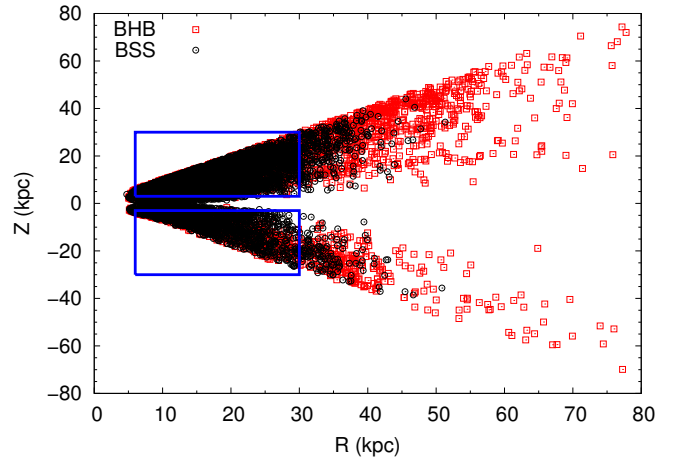}
\caption{Distribution of height above the Galactic plane, Z, as a
function of Galactocentric distance, R, for the BSSs (black circles)
and BHB stars (red squares) in our sample. The blue rectangles limit the 
regions where we have obtained the relative stellar counts, avoiding the 
difficulties (under/over counts of BSSs and BHB stars at the bright/faint 
limits of our sample), as noted in the text: 3 kpc $<$ $|$Z$|$ $<$ 30 
kpc and 6 kpc $<$ R $<$ 30 kpc. \\} 
\label{distance_R_Z_all}
\end{figure}

\begin{figure} 
\epsscale{1.22} 
\plotone{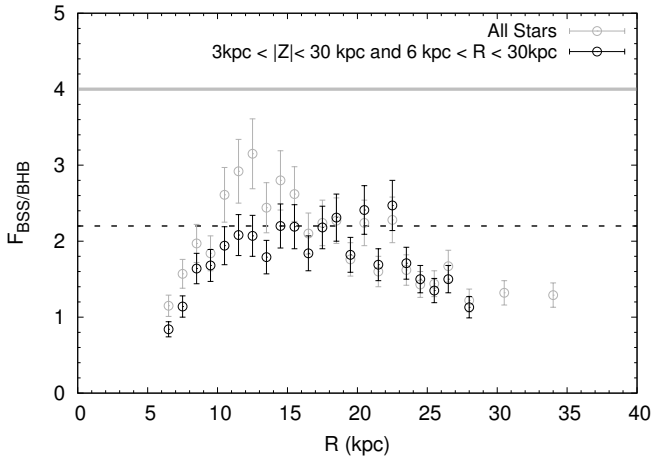}
\caption{Comparison of the reative frequency, F$_{\rm BSS/BHB}$, as a function of 
the Galactocentric distance, for all sample (gray circles) and considering the 
limited regions defined in Figure~\ref{distance_R_Z_all} (black circles). 
The gray line represents the value of F$_{\rm BSS/BHB}$ estimated in the Solar 
Neighborhood by \citet{Preston94} and the dashed line represents the value of 
F$_{\rm BSS/BHB}$ found in nearby dwarf galaxies by \citet{Momany07}. The 
F$_{\rm BSS/BHB}$ found for the Galactic halo system are more consistent with the 
value of F$_{\rm BSS/BHB}$ for nearby dwarf galaxies than to that found for the 
Solar Neighborhood. \\}
\label{fbsxbhb_R_limited}
\end{figure}

The frequencies presented in Figure~\ref{fbsxbhb_Rerrors} were derived
to compare the differences resulting from the adoption of different
absolute magnitude calibrations. However, as foreshadowed above, we
cannot evaluate the value of F$_{\rm BSS/BHB}$ in the field without taking
into account that there exist regions of the sampled volume that are not
simultaneously occupied by BSSs and BHB stars. Near the bright magnitude
limit of the SDSS scans ($g~\sim~14.5$) the closest selected A-type
stars will preferably be BSSs, as BHB stars at similar distances will
generally be saturated. Indeed, at $|$Z$|$ $<$ 1 kpc, our sample
includes more than 250 BSSs and only 4 BHB stars. To be safe, we have
obtained our counts avoiding the region within $|$Z$|$ $=$ 3 kpc. At
larger distances ($|$Z$|$ $>$ 30 kpc and R $>$ 30 kpc), our sample
preferentially includes BHB stars, and only a few BSSs.
Figure~\ref{distance_R_Z_all} summarizes the restrictions we have adopted to
derive the relative frequencies, F$_{\rm BSS/BHB}$, shown in 
Figure~\ref{fbsxbhb_R_limited}.

From inspection of Figure~\ref{fbsxbhb_R_limited}, the relative blue-straggler
frequency in the Galactic halo system is consistent with the value of
F$_{\rm BSS/BHB}$ obtained for the nearby dwarf galaxies, but not with
that obtained for the Solar Neighborhood. The average value of F$_{\rm
BSS/BHB}$ from R $=$ 7 kpc to R $=$ 27 kpc is 1.83 $\pm$ 0.23.

\begin{figure} 
\epsscale{1.22} 
\plotone{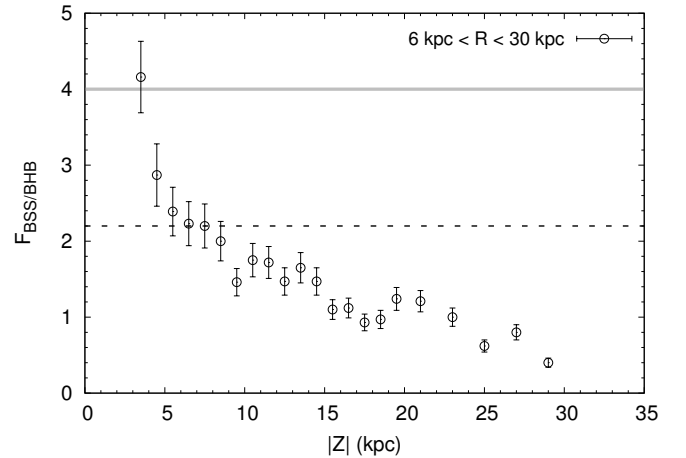}
\caption{Comparison of F$_{\rm BSS/BHB}$, as a function of the 
distance from the Galactic plane, $|$Z$|$, considering the limited regions 
defined in Figure~\ref{distance_R_Z_all}. The gray line represents 
the value of F$_{\rm BSS/BHB}$ estimated in the Solar Neighborhood by 
\citet{Preston94} and the dashed line represents the value of F$_{\rm BSS/BHB}$ 
found in nearby dwarf galaxies by \citet{Momany07}. The relative 
frequency of BSSs rises closer to the Galactic plane, in agreement
with the Solar Neighborhood frequency obtained by \citet{Preston94}. \\}
\label{fbsxbhb_Zmod_limited}
\end{figure}

\citet{Preston94} evaluated the F$_{\rm BSS/BHB}$ in the Solar Neighborhood 
using stars within 2 kpc of the Sun, a region our data precludes us from
exploring. However, we can also examine the F$_{\rm BSS/BHB}$ values from
another perspective, evaluating the change in frequency with distance
from the Galactic plane, $|$Z$|$, from $|$Z$|$ $=$ 3 kpc to $|$Z$|$ $=$
30 kpc, still considering R between 6 and 30 kpc, as shown in
Figure~\ref{fbsxbhb_Zmod_limited}. From inspection of this Figure, the
relative frequency of BSSs increases at distances closer to the Galactic
plane, reaching a maximum F$_{\rm BSS/BHB}$ $=$4.11 $\pm$ 0.47 at 3 kpc
$<$ $|$Z$|$ $<$ 4 kpc, in agreement with the previous calculation of the
Solar Neighborhood frequency.

\section{Possible Impact of the Presence of Streams}
\label{c5}

It has been shown by a number of recent investigations 
\citep[e.g.,][]{Majewski03,Belokurov06,Koposov12} that
the relative importance of debris streams increases with distance from
the Galactic center, hence we wish to explore the possible impact of
this on our results, making use of our observed stellar radial
velocities.

\subsection{Distribution of Galactocentric Radial Velocities}

Due to their relatively low spatial density, BSSs having R $>$ 25 kpc
cannot be readily distinguished as members of stellar streams in the
outer region of the Galactic halo based on their distances alone.
Analysis of their radial velocities can, however, be used to check for
the possible presence of ``extragalactic'' blue stragglers (i.e., those
captured from dissolved satellites) among the stars in the field
population.

In order to carry out this analysis, the heliocentric radial velocities
(V$_{\rm HRV}$) of BSSs were converted to the Galactic standard of rest
(GSR) frame. The adopted velocity for the Local Standard of Rest
(V$_{\rm LSR}$) is 220 km s$^{-1}$, and a solar motion ($U,V,W$) = (+10.1,+4.0,+6.7)
km s$^{-1}$, defined in a right-handed Galactic system
with $U$ pointing towards the Galactic center, $V$ in the direction of
Galactic rotation, and $W$ towards the north Galactic pole \citep{Hogg05}.
Hereafter, V$_{\rm GSR}$ is the radial velocity in the GSR frame (i.e.,
the radial velocity component along the star-Sun direction, corrected
for Galactic rotation). Then,

\begin{eqnarray}
{\rm {V_{GSR}}} = {\rm {V_{HRV}}} + ({\rm {V_{LSR}}} + 5.2)\sin{l}\cos{b} \nonumber \\
 + 10\cos{l}\cos{b} + 7.2\sin{b}.
\label{vg}
\end{eqnarray}

Using the V$_{\rm GSR}$ values, one can search for stellar streams
associated with the various structural components of the Galaxy,
including the thick disk, the inner halo, and the outer halo. By
consideration of the different distributions, segregated by distance
from the Galactic plane, $|$Z$|$, the inner-halo component of the Milky
Way is expected to dominate the population of halo stars found at
distances up to $|$Z$|$ $\sim$10-15 kpc. The outer-halo component is
expected to dominate in regions beyond $|$Z$|$ $\sim$15-20 kpc
\citep{Carollo07,Carollo10,Beers12}. It is important to note that the limits used to
separate these in the previous work considered ${\rm {Z_{max}}}$ and not
$|$Z$|$. The derived parameter ${\rm {Z_{max}}}$ depends on the adopted
gravitational potential. In this work we assume that, if there are BSSs
whose origin is extragalactic, differences in their
radial-velocity distributions in the Galactic standard of rest will be
noticed in intervals of either ${\rm {Z_{max}}}$ or $|$Z$|$.

\begin{figure} 
\epsscale{1.22} 
\plotone{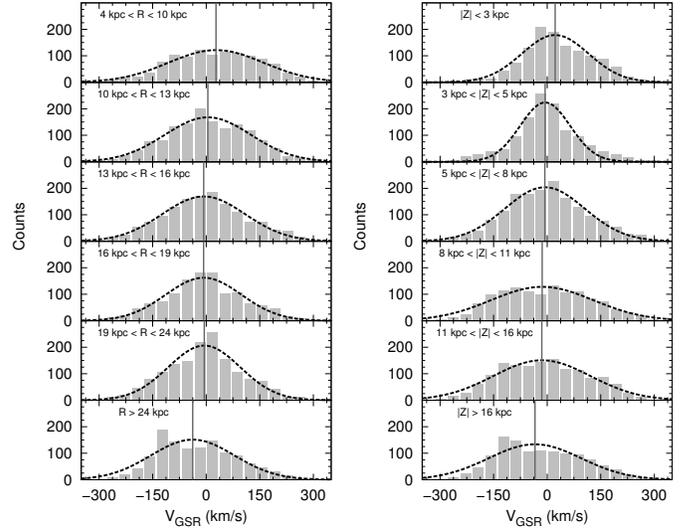}
\caption{Distribution of V$_{\rm GSR}$ for BSSs, considering the regions limited
by R (left panels) and $|$Z$|$ (right panels). The main distribution found in
each panel (usually with mean value close to zero and dispersion of
about 100-150 km s$^{-1}$) represents typical values of radial
velocities for halo stars. The vertical lines show the mean value of V$_{\rm GSR}$
for each single Gaussian distribution fitted (dashed black line). It is evident 
that there is a shift in the mean value of the V$_{\rm GSR}$ when we reach the 
outer-halo region seen in the bottom panels. \\}
\label{vgsr_RZ}
\end{figure}

\begin{figure} 
\epsscale{1.22} 
\plotone{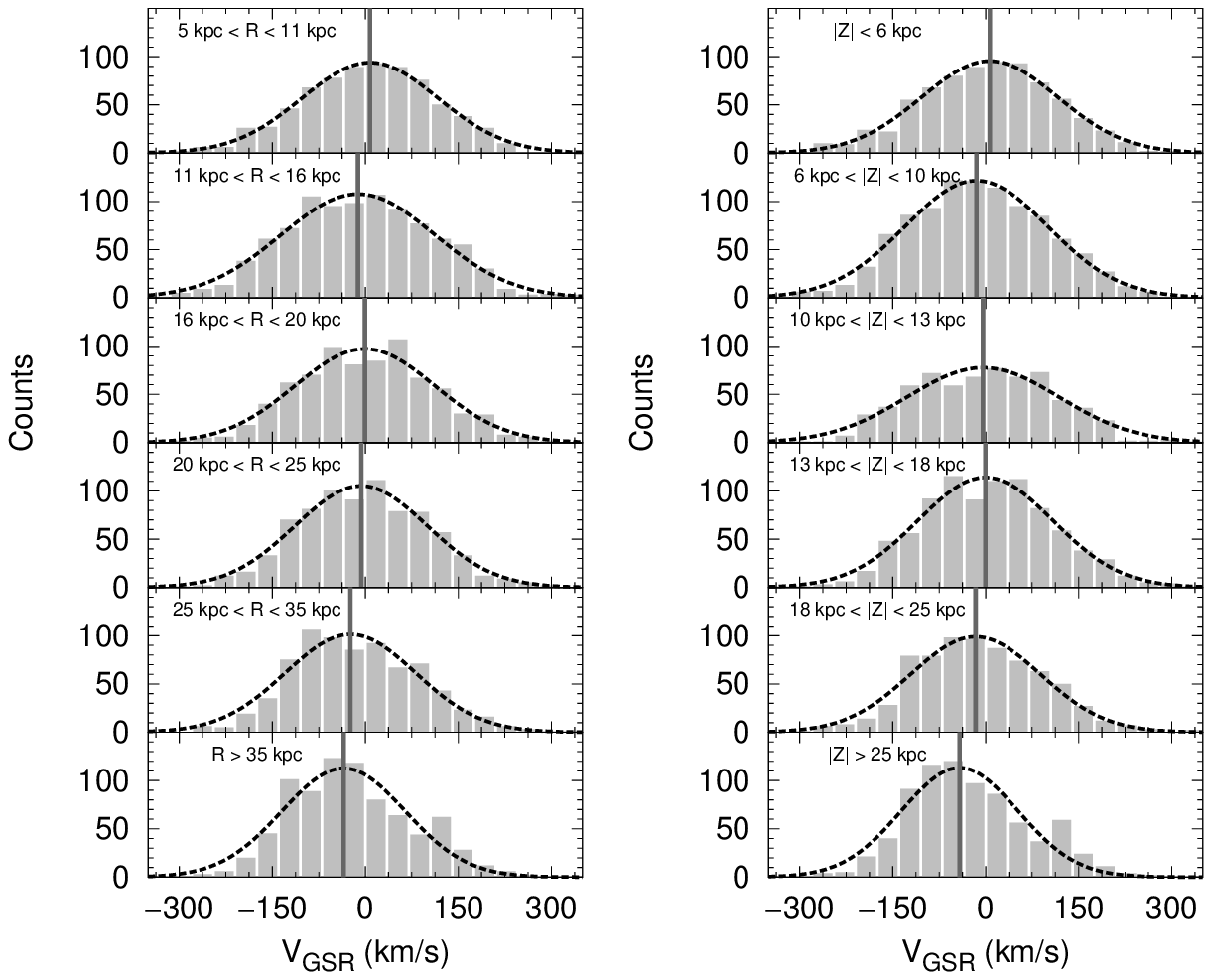}
\caption{Distribution of V$_{\rm GSR}$ for BHBs, considering the regions limited
by R (left panels) and $|$Z$|$ (right panels). The main distribution found in
each panel (usually with mean value close to zero and dispersion of
about 100-150 km s$^{-1}$) represents typical values of radial
velocities for halo stars. The vertical lines show the mean value of V$_{\rm GSR}$
for each single Gaussian distribution fitted (dashed black line). It is evident 
that there is a shift in the mean value of the V$_{\rm GSR}$ when we reach the 
outer-halo region seen in the bottom panels. \\}
\label{vgsr_RZ_BHB}
\end{figure}

\begin{figure} 
\epsscale{1.22} 
\plotone{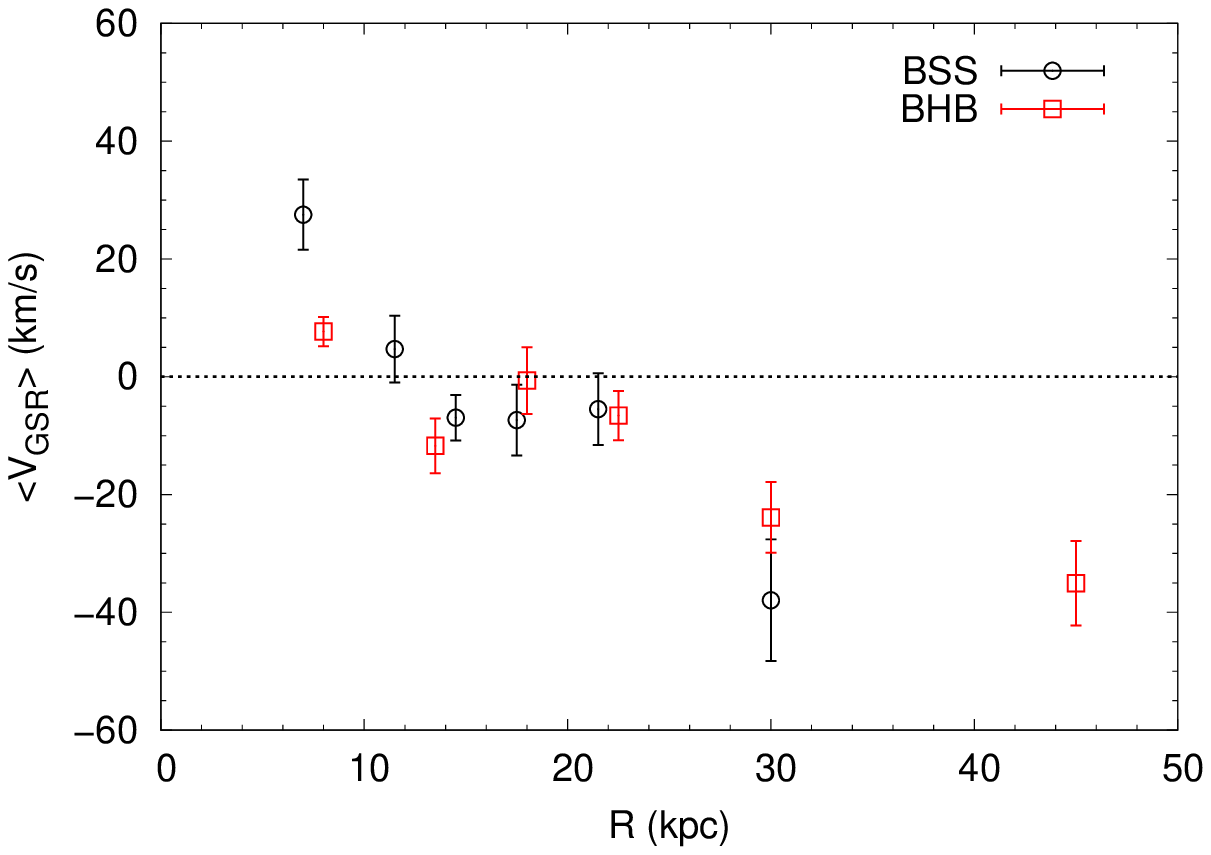}
\caption{Mean values of V$_{\rm GSR}$ shown in Figures~\ref{vgsr_RZ} 
and \ref{vgsr_RZ_BHB}, as a function of R. The black circles are the
BSSs and the red are squares the BHBs. Error bars were estimated using 
the standard deviations of the fitted Gaussians. \\}
\label{Shift_R}
\end{figure}

\begin{figure} 
\epsscale{1.22} 
\plotone{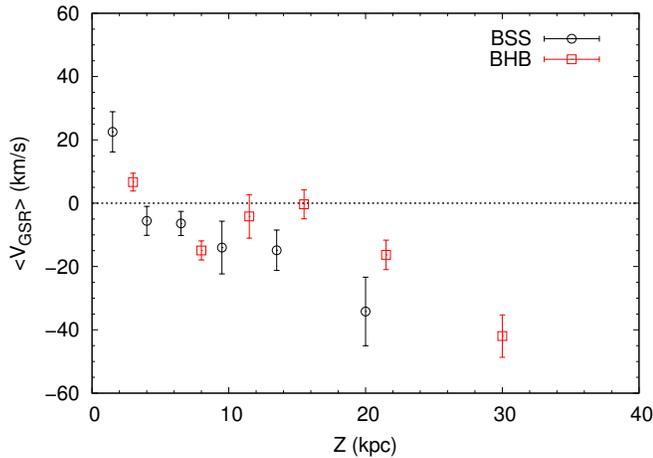}
\caption{Mean values of V$_{\rm GSR}$ shown in Figures~\ref{vgsr_RZ} 
and \ref{vgsr_RZ_BHB}, as a function of $|$Z$|$. The black circles are
the BSSs and the red squares are the BHBs. Error bars were estimated
using the standard deviations of the fitted Gaussians. \\}
\label{Shift_Z}
\end{figure}

The velocity distributions of BSSs and BHB stars, considering the
approximate limits where each Galactic component becomes dominant, are
shown in Figures~\ref{vgsr_RZ} and \ref{vgsr_RZ_BHB}, respectively. The
mean value for the distributions is usually close to zero (with
dispersion of about 100-150 km s$^{-1}$), representing the typical values of
radial velocities for halo stars \citep{Hogg05}. The mean value of
velocities for the BSSs and BHB stars for the most distant stars is
shifted toward negative values. This contribution is likely due to a stellar 
stream, such as Sagittarius, in both the distance from the Galactic center (R -
left panels) and in distance from the Galactic plane ($|$Z$|$ - right
panels). The dashed lines represent Gaussian fits of these
distributions, while the vertical lines indicate their mean values,
which are always negative for the most distant stars. These behaviors
are evident from inspection of Figures~\ref{Shift_R} and
\ref{Shift_Z}.

\subsection{Comparison with the Sagittarius Stream}

\begin{figure} 
\epsscale{1.22} 
\plotone{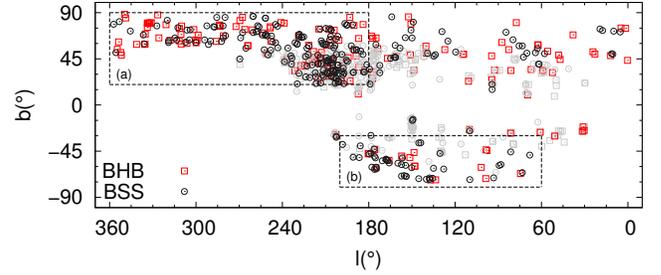}
\caption{Galactic celestial coordinates of BSSs at $|$Z$|$ $>$ 17 kpc
or $|$Z$|$ $<$ 17 kpc and R $>$ 24 kpc. The BSSs 
(black circles) and BHB stars (red squares) with $-$200 km s$^{-1}$ $<$ 
V$_{\rm GSR}$ $<$ $-$50 km s$^{-1}$ are preferably located over similar Galactic 
coordinates as the Sagittarius Stream. The large dashed rectangles highlight 
the regions where the stream is dominant in (a) The Northern Galactic 
Hemisphere, and in (b) the Southern Galactic Hemisphere. \\}
\label{lxb_stream}
\end{figure}

We now compare the V$_{\rm GSR}$ for BSSs and BHB stars found at $|$Z$|$
$>$ 17 kpc with typical values for the Sagittarius Stream
\citep{Majewski03}, in order to verify the hypothesis that at least a
portion of the BSSs and BHB stars in our sample can be associated with
this stream, shown in Figure~\ref{lxb_stream}. This Figure highlights
the positions (in Galactic coordinates) of BSSs (black circles) and BHB
stars (red squares) having velocities in the range $-$200 km s$^{-1}$ $<$
V$_{\rm GSR}$ $<$ $-$50 km s$^{-1}$ (the typical Sagittarius Stream
range of velocities). From inspection of this Figure, these stars are
preferably located over the same coordinates as the Sagittarius Stream
\citep{Belokurov06,Koposov12}, highlighted by the large dashed
rectangles: (a) In the Northern Galactic Hemisphere, and in (b) the
Southern Galactic Hemisphere. The mean heliocentric distance
distribution found for both the BSSs and BHBs with these velocities is
D$_{{\rm {Sg}}}$ $=$ 21.9 $\pm$ 2.2 kpc. Within 2$\sigma$, these
distances agree with the distance to the Sagittarius Stream measured by
\citet{Watkins09} using RR-Lyr$\ae{}$ stars: D$_{{\rm {Sg}}}$ $=$ 26.1
$\pm$ 5.6 kpc. We conclude that at least a portion of the field BSSs and
BHB stars are likely to have originated from the Sagittarius Stream,
i.e., they are of extragalactic origin.

\begin{figure}
\epsscale{1.22}
\plotone{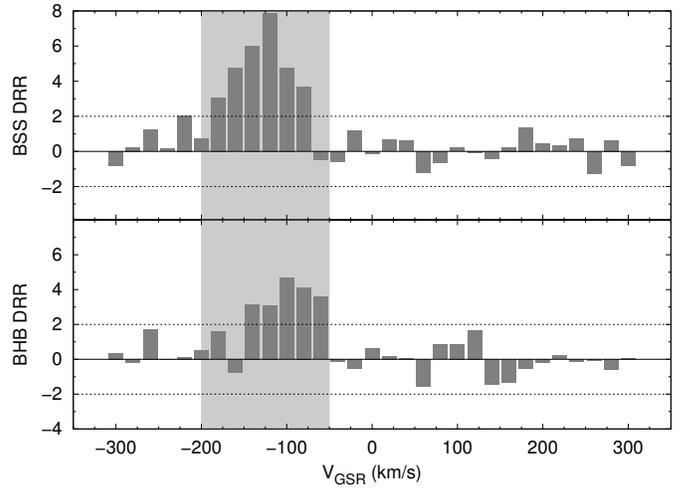}
\caption{Double Residual Root (DRR) plot for the Gaussian fits of the
distribution of velocities (V$_{\rm GSR}$) shown inside the blue rectangles 
of Figure~\ref{distance_R_Z_all}, limited by $|$Z$|$ $>$ 17 kpc. 
The top panel shows the BSSs whose velocities suggest association with the 
Sagittarius Stream (gray shaded region). The bottom panel shows the same 
distribution for the BHB stars. The dashed lines at $\pm$ 2 indicate an 
approximate 95\% significance level. \\}
\label{DRR}
\end{figure}

Figure~\ref{DRR} shows the excess counts of stars relative to the
Gaussian fits of the velocity distributions (V$_{\rm GSR}$) for the
BSSs and BHB stars in the blue rectangles of Figure~\ref{distance_R_Z_all}, 
limited by $|$Z$|$ $>$ 17 kpc. We have employed a variation of a simple 
difference plot, showing the distribution of the so-called Double Root Residuals 
\citep[DRRs; see][]{Gebhardt91}. The DRR technique takes advantage of the 
variance-stabilizing properties of a square-root transformation. A DRR plot 
graphically emphasizes {\it where} in a distribution there is significant lack 
of fit between the data and model -- the square-root transformation puts the 
residuals throughout the fit on an equal footing. If the model is an adequate 
fit to the data, then the DRRs are roughly equivalent to normal deviates. Thus, 
a DRR with numerical value exceeding $\pm 2$ is significant at the 95\% 
(2$\sigma$) level, whereas DRRs with absolute magnitude less than 1 indicate a
reasonable agreement between data and model.

The top panel of Figure~\ref{DRR} shows the DRRs for BSSs with
velocities suggesting association with the Sagittarius Stream (gray
shaded region), while the bottom panel shows the same restrictions for
BHB stars. There exists an excess of about $\sim$400 BSSs and $\sim$250
BHB stars with values of V$_{\rm GSR}$ similar to the Sagittarius
Stream. This leads to a minimum contamination due to extragalactic BSSs
arising from the Sagittarius Stream of about $5\%$ of the total BSS
sample. The contamination for BHB stars is even higher, reaching at
least $20\%$. Applying the limits described on Figure~\ref{distance_R_Z_all} 
and the restrictions of DRR analysis, we found an excess of 246 BSSs and 
146 BHB stars in the Sagittarius Stream region. Thus, an estimate of 
F$_{\rm BSS/BHB}$ for the Sagittarius Stream is 1.68 $\pm$ 0.21, in 
agreement with the mean value of the relative blue-straggler frequency 
in the Galactic halo and in the nearby dwarf galaxies within 3$\sigma$. 
We also verified that the impact of these likely members of the Sagittarius 
Stream do not change the F$_{\rm BSS/BHB}$ values derived by our analysis.

\section{Discussion and Conclusions}
\label{c6}

We have identified 8001 BSSs and 4796 BHB stars from the SDSS/SEGUE
spectroscopic database, obtained through DR8. We find that the BSSs
become more frequent than BHB stars for $g > 17$, and that the mean
value for the blue-straggler frequency in the Galactic halo is 1.83
$\pm$ 0.23, which agrees well with the expected upper limit for BSSs in
the nearby dwarf galaxies \citep[$\sim$ 2.2]{Momany07} within 2$\sigma$.
This quantity may be an important input for simulations of stellar
populations in the Galaxy, in particular for studies that include BSSs
in these calculations \citep{Chen09,Conroy10, Xin11, Zhang12}. We also
verified that, for the region closest to the Galactic plane that we can
study (3 kpc $<$ $|$Z$|$ $<$ 4 kpc), the relative frequency of blue
stragglers, F$_{\rm BSS/BHB}$ = 4.11 $\pm$ 0.47, agrees well with
\citet{Preston94} BSSs frequency estimate in the Solar Neighborhood.
The relative frequency drops with distance from the plane, reaching
$\sim~1.5-2.0$ in the inner-halo region; this ratio continues to decline
to $\sim~$1.0 in the outer-halo region.

Another interesting result from our analysis is the verification of the
extragalactic origin of a small fraction ($\sim 5$\%) of BSSs (a
somewhat higher fraction, $\sim 20$\%, is obtained for BHB stars) found
among stars in the outer-halo region. 

The distribution of BSSs throughout the Galaxy can be used to constrain
how the frequency of primordial binary systems (assuming this is the
primary formation channel) varies with distance from the Galactic
center. Our analysis indicates that the relative number of BSSs
decreases with both R and $|$Z$|$, thus it would follow that the
relative frequency of primordial binary systems should decrease as well.

\acknowledgments R.M.S., S.R. and H.M.R. acknowledge CAPES (PROEX), CNPq, 
PRPG/USP, FAPESP and INCT-A funding. V.M.P. acknowledges support from 
the Gemini Observatory. T.C.B. acknowledges partial support from grants
PHY 08-22648; Physics Frontier Center/JINA, and PHY 14-30152; JINA
Center for the Evolution of the Elements (JINA-CEE),
awarded by the US National Science Foundation. X.$-$X. Xue acknowledges
the Alexandre Von Humboldt foundation for a fellowship, the DFG's
SFB-881 grant ``The Milky Way System'', and the National Natural Science
Foundation of China under grant Nos. 11103031, 11233004, 11390371.
The authors would also like to thank the referee, George W. Preston, 
for his useful comments, and for inspiring this study with his previous 
work on blue-straggler stars.

\bibliography{bibliografia} \bibliographystyle{apj}

\end{document}